\documentclass[%
 aip,
rsi,%
citeautoscript,
 amsmath,amssymb,
preprint,%
]{revtex4-1}

\usepackage{graphicx}
\usepackage{dcolumn}
\usepackage{bm}
\usepackage[colorlinks=true,linkcolor=blue,citecolor=cyan]{hyperref}
\usepackage[version=4]{mhchem}

\usepackage{color}
\usepackage[dvipsnames]{xcolor}
\usepackage{multirow}
\usepackage{rotating}
\usepackage{threeparttable}
\usepackage{braket}
\usepackage{bbold}
\usepackage{inputenc}
\begin{document} 

\title{The DIRAC code for relativistic molecular calculations}

\author{Trond Saue}
\email{trond.saue@irsamc.ups-tlse.fr}
\homepage{http://dirac.ups-tlse.fr/saue}
\affiliation{Laboratoire de Chimie et Physique Quantique, UMR 5626 CNRS
--- Universit{\'e} Toulouse III-Paul Sabatier, 118 route de Narbonne,
F-31062 Toulouse, France}

\author{Radovan Bast}
\email{radovan.bast@uit.no}
\homepage{https://bast.fr}
\affiliation{Department of Information Technology, 
UiT The Arctic University of Norway, 
N--9037 Troms{\o}, Norway}

\author{Andr{\'e} Severo Pereira Gomes}
\email{andre.gomes@univ-lille.fr}
\affiliation{Universit\'e de Lille, CNRS, 
UMR 8523 -- PhLAM -- Physique des Lasers, 
Atomes et Mol\'ecules, F-59000 Lille, France}

\author{Hans J{\o}rgen {Aa}. Jensen}
\email{hjj@sdu.dk}
\affiliation{Department of Physics, Chemistry and Pharmacy,
University of Southern Denmark,
DK-5230 Odense M, Denmark}

\author{Lucas Visscher}
\email{l.visscher@vu.nl}
\affiliation{Department of Chemistry and Pharmaceutical Sciences,
Vrije Universiteit Amsterdam,
NL-1081HV Amsterdam, The Netherlands}

\author{Ignacio Agust{\'{\i}}n Aucar}
\email{agustin.aucar@conicet.gov.ar}
\affiliation{Instituto de Modelado e Innovaci\'on Tecnol\'ogica, CONICET, and Departamento de F\'{\i}sica - Facultad de Ciencias Exactas y Naturales, UNNE, Avda.~Libertad 5460, W3404AAS, Corrientes, Argentina}

\author{Roberto Di Remigio}
\affiliation{Hylleraas Centre for Quantum Molecular Sciences, Department of Chemistry, UiT The Arctic University of Norway, N-9037 Troms{\o}, Norway}

\author{Kenneth G. Dyall}
\email{diracsolutions@gmail.com}
\affiliation{Dirac Solutions, 10527 NW Lost Park Drive, Portland, OR 97229, U.S.A.}

\author{Ephraim Eliav}
\email{ephraim@tau.ac.il}
\affiliation{School of Chemistry, Tel Aviv University, Ramat Aviv, Tel Aviv, 69978, Israel}

\author{Elke Fasshauer}
\email{elke.fasshauer@gmail.com}
\affiliation{Department of Physics and Astronomy, Aarhus University, Ny Munkegade 120, 8000 Aarhus, Denmark}

\author{Timo Fleig}
\email{timo.fleig@irsamc.ups-tlse.fr}
\homepage{http://dirac.ups-tlse.fr/fleig}
\affiliation{Laboratoire de Chimie et Physique Quantique, UMR 5626 CNRS
--- Universit{\'e} Toulouse III-Paul Sabatier, 118 route de Narbonne,
F-31062 Toulouse, France}

\author{Lo\"ic Halbert}
\affiliation{Universit\'e de Lille, CNRS, 
UMR 8523 -- PhLAM -- Physique des Lasers, 
Atomes et Mol\'ecules, F-59000 Lille, France}

\author{Erik Donovan Hedeg{\aa}rd}
\affiliation{Division of Theoretical Chemistry, Lund University, Chemical Centre, P. O. Box 124, SE-221 00 Lund, Sweden}

\author{Benjamin Helmich-Paris}
\affiliation{Max-Planck-Institut f{\"u}r Kohlenforschung, Kaiser-Wilhelm-Platz 1, 45470 M{\"u}lheim an der Ruhr, Germany}

\author{Miroslav Ilia\v{s}}
\email{Miroslav.Ilias@umb.sk}
\affiliation{Department of Chemistry, Faculty of Natural Sciences, Matej Bel University, Tajovsk{\'e}ho 40, 974 01 Bansk{\'a} Bystrica, Slovakia}

\author{Christoph Jacob}
\email{c.jacob@tu-braunschweig.de}
\affiliation{Technische Universit{\"a}t Braunschweig, Institute of Physical and Theoretical Chemistry, Gau{\ss}str.~17, 38106 Braunschweig, Germany}

\author{Stefan {Knecht}}
\email{stefan.knecht@phys.chem.ethz.ch}
\affiliation{ETH Z{\"u}rich, Laboratorium f\"ur Physikalische Chemie, Vladimir-Prelog-Weg 2, 8093 Z{\"u}rich, Switzerland}

\author{Jon K. Laerdahl}
\affiliation{Department of Microbiology, Oslo University Hospital, Oslo, Norway}

\author{Marta {Lopez Vidal}}
\affiliation{Department of Chemistry, Technical University of Denmark, 2800 Kgs. Lyngby, Denmark}

\author{Malaya K. Nayak}
\email{mknayak@barc.gov.in, mk.nayak72@gmail.com}
\affiliation{Theoretical Chemistry Section, Bhabha Atomic Research Centre, Trombay, Mumbai - 400085, India}

\author{Ma{\l}gorzata Olejniczak}
\email{malgorzata.olejniczak@cent.uw.edu.pl}
\affiliation{Centre of New Technologies, University of Warsaw, S. Banacha 2c, 02-097 Warsaw, Poland}

\author{J{\'o}gvan Magnus Haugaard Olsen}
\affiliation{Hylleraas Centre for Quantum Molecular Sciences, Department of Chemistry, UiT The Arctic University of Norway, N-9037 Troms{\o}, Norway}

\author{Markus Pernpointner}
\email{markpp@gmx.de}
\affiliation{Kybeidos GmbH, Heinrich-Fuchs-Str. 94, 69126 Heidelberg, Germany}

\author{Bruno Senjean}
\email{bsenjean@gmail.com}
\affiliation{Instituut-Lorentz, Universiteit Leiden, P.O. Box 9506, 2300 RA Leiden, The Netherlands}
\affiliation{Department of Chemistry and Pharmaceutical Sciences,
Vrije Universiteit Amsterdam,
NL-1081HV Amsterdam, The Netherlands}

\author{Avijit Shee}
\email{ashee@umich.edu}
\affiliation{Department of Chemistry, University of Michigan, Ann Arbor, Michigan 48109,USA}

\author{Ayaki Sunaga}
\email{sunagaayaki@gmail.com}
\affiliation{Department of Chemistry, Tokyo Metropolitan University, 1-1 Minami-Osawa, Hachioji-city, Tokyo 192-0397,Japan}

\author{Joost N. P. \surname{van Stralen}}
\email{joost.vanstralen@tno.nl}
\affiliation{Department of Chemistry and Pharmaceutical Sciences,
Vrije Universiteit Amsterdam,
NL-1081HV Amsterdam, The Netherlands, Current address: TNO, Energy Transition Studies, Radarweg 60, NL-1043NT, Amsterdam, The Netherlands}

\date{\today}

\begin{abstract}
DIRAC is a freely distributed general-purpose program system for 1-, 2- and 4-component relativistic molecular calculations at the level of Hartree--Fock, Kohn--Sham (including range-separated theory), multiconfigurational self-consistent-field, multireference configuration interaction, coupled cluster and electron propagator theory. At the self-consistent-field level a highly original scheme, based on quaternion algebra, is implemented for the treatment of both spatial and time reversal symmetry. DIRAC features a very general module for the calculation of molecular properties that to a large extent may be defined by the user and further analyzed through a powerful visualization module. It allows the inclusion of environmental effects through three different classes of increasingly sophisticated embedding approaches:  the  implicit  solvation  polarizable  continuum  model,  the  explicit polarizable embedding, and frozen density embedding models. DIRAC was one of the earliest codes for relativistic molecular calculations and remains a reference in its field.
\end{abstract}


\maketitle

\section{Introduction}
DIRAC is a general-purpose program system for relativistic molecular calculations and is named in honor of P.~A.~M. Dirac (\textbf{P}rogram for \textbf{A}tomic and \textbf{M}olecular \ \textbf{D}irect \textbf{I}terative \textbf{R}elativistic \textbf{A}ll--electron \textbf{C}alculations), who formulated\cite{Dirac:PRSLA1928} his celebrated relativistic wave equation for the electron in 1928.
The beginnings of the DIRAC code can be traced back to the 4-component relativistic Hartree--Fock code written by Trond Saue during his Master thesis, defended at the University of Oslo, Norway, in 1991. The original code stored all integrals, provided by the HERMIT code,\cite{prog:hermit} on disk, but during Saue's Ph.D. thesis, defended in 1996, the code was extended to direct Self-Consistent Field (SCF) with integral screening\cite{Saue:MP1997} and a highly original symmetry scheme based on quaternion algebra.\cite{Saue:Jensen:JCP1999} A postdoctoral stay in  1996-97 with Hans J{\o}rgen  Aagaard Jensen at the University of Southern Denmark focused on molecular properties, with the implementation of the calculation of expectation values and linear response functions\cite{Saue:Jensen:JCP2003} at the SCF level. Lucas Visscher, who had written a 4-component direct Restricted Active Space (RAS) Configuration Interaction (CI) code for the MOLFDIR program system\cite{prog:molfdir} during his Ph.D. thesis, defended at the University of Groningen in 1993, did a postdoctoral stay with Jens Oddershede in Odense during the years 1996-97 and joined forces and code with Jensen and Saue to create the DIRAC program system. Since then the main author team has been joined by Radovan Bast and Andre Severo Pereira Gomes in addition to almost fifty contributors, and Odense has since 1997 hosted an annual ``family'' meeting for DIRAC developers. In addition to the above authors, we would like to highlight the contributions to code infrastructure by J{\o}rn Thyssen and Miroslav Ilia{\v{s}}. The latest version of the code, DIRAC19\cite{DIRAC19}, was released December 12, 2019.

In the next section we give a brief overview of the DIRAC program. Then, in section \ref{sec:impdet}, we provide some implementation details, with focus on features that are little documented elsewhere and/or may be a source of confusion for DIRAC users.

\section{Program Overview}

\subsection{Hamiltonians}
Within the Born--Oppenheimer approximation and using the atomic units system, the electronic Hamiltonian may be expressed as
\begin{equation}
\hat{H}=V_{NN}+\sum_i\hat{h}(i)+\frac{1}{2}\sum_{i\ne j}\hat{g}(i,j);\quad V_{NN}=\frac{1}{2}\sum_{A\ne B}\frac{Z_AZ_B}{R_{AB}},
\end{equation}
where $V_{NN}$ represents the repulsion energy arising from point nuclei fixed in space. 
Notwithstanding the challenges associated with
specific choices of the one-and two-electron operators $\hat{h}(i)$ and $\hat{g}(i,j)$, most quantum chemical methods can be formulated just from this generic form. This becomes perhaps even more evident
by considering the electronic Hamiltonian in second quantized form
\begin{equation}\label{eq:secham}
  \hat{H}=V_{NN}+\sum_{pq}h^p_qa^{\dagger}_pa^{\phantom{\dagger}}_q+\frac{1}{4}\sum_{pqrs}V^{pr}_{qs}a^{\dagger}_pa^{\dagger}_ra^{\phantom{\dagger}}_sa^{\phantom{\dagger}}_q;
  \quad\begin{array}{lcl}h^p_q&=&\langle p|\hat{h}|q\rangle\\V^{pr}_{qs}&=&\langle pr|\hat{g}|qs\rangle-\langle pr|\hat{g}|sq\rangle.
  \end{array}
\end{equation}
In this form, practical for actual implementations, the Hamiltonian is given by strings of creation and annihilation operators combined with one- and two-electron integrals.
In relativistic calculations the integrals are generally complex, in contrast to the nonrelativistic domain, and contain fewer zero elements, since spin symmetry is lost.

The DIRAC code features several electronic Hamiltonians, allowing molecular electronic structure calculations at the 4-, 2- and 1-component level.
4-component relativistic calculations are sometimes referred to as `fully relativistic' in contrast to `quasirelativistic' 2-component calculations.
However, a fully relativistic two-electron interaction, which would contain magnetic interactions and effects of retardation in addition to electrostatics, 
is not readily available in closed form, rendering this terminology somewhat misleading.

The default Hamiltonian of DIRAC is the 4-component Dirac--Coulomb Hamiltonian, using the Simple Coulombic correction,\cite{Visscher:tca1997} 
which replaces the expensive calculation of two-electron integrals over small component basis functions by an energy correction.
The one-electron part is the Hamiltonian $\hat{h}_D$ of the time-independent Dirac equation in the molecular field, that is,
the field of nuclei fixed in space
\begin{equation}
  \hat{h}_D\psi=\left[\begin{array}{ll}V_{eN}&c(\boldsymbol{\sigma}\cdot\mathbf{p})\\
      c(\boldsymbol{\sigma}\cdot\mathbf{p})&V_{eN}-2m_ec^2\end{array}\right]
  \left[\begin{array}{c}\psi^L\\\psi^S\end{array}\right]=\left[\begin{array}{c}\psi^L\\\psi^S\end{array}\right]E;\quad
  \begin{array}{lcl}
  V_{eN}(\mathbf{r})=\sum_A\frac{-e}{4\pi\varepsilon_0} \int_{\mathbb{R}^3}\frac{\rho_A(\mathbf{r}^\prime)}{|\mathbf{r}^\prime-\mathbf{r}|}d^3\mathbf{r}^\prime\\
  \int_{\mathbb{R}^3} \rho_A(\mathbf{r})d^3\mathbf{r}=Z_Ae \end{array},
\end{equation}
where $c$ is the speed of light, $\boldsymbol{\sigma}$ the vector of Pauli spin matrices, $\mathbf{p}$ the momentum operator and $V_{eN}$ the electron--nucleus interaction. 
The default model of the nuclear charge distribution is the Gaussian approximation,\cite{Visscher:adndt1997} but a point nucleus model is also available.
The default two-electron operator of DIRAC is the instantaneous Coulomb interaction
\begin{equation}
  \hat{g}^C(1,2)=\frac{1}{r_{12}},
\end{equation}
which constitutes the zeroth-order term and hence the nonrelativistic limit\cite{saue:aqc2005} of an expansion in $c^{-2}$ of the fully relativistic two-electron interaction in the Coulomb gauge. 
It should be noted, though, that the presence of $\hat{g}^C(1,2)$ induces spin--same orbit (SSO) interaction, just as the presence of $V_{eN}$ induces spin-orbit interaction associated with the
relative motion of nuclei with respect to electrons.\cite{saue:cpc2011} Spin-other-orbit (SOO) interaction may be included by adding the Gaunt term, which is available at the SCF level.
Spin-orbit interaction may be eliminated by transforming to the modified Dirac equation and removing spin-dependent terms.\cite{dyall:jcp1994} In the quaternion formulation of SCF calculations in DIRAC this
corresponds to removing the quaternion imaginary parts of the Fock matrix.\cite{visscher:jcp2000}
It is also possible to carry out 4-component \textit{nonrelativistic} calculations using the L{\'e}vy--Leblond Hamiltonian.\cite{levy-leblond:cmp1967,visscher:jcp2000} With respect to the Schr{\"o}dinger Hamiltonian, equivalent within kinetic balance
and also available in DIRAC, the L{\'e}vy--Leblond Hamiltonian has advantages in the calculation of magnetic properties, since it is linear in vector potentials.

A troublesome aspect of the Dirac Hamiltonian is the presence of solutions of negative energy. Over the years, there has been extensive work on eliminating the positronic degrees
of freedom of the Dirac Hamiltonian, leading to approximate 2-component Hamiltonians, such as the Douglas--Kroll--Hess (DKH)\cite{douglas:kroll:ap1974,hess:pra1985,hess:pra1986} and zeroth-order regular approximation (ZORA)\cite{durand:ps1986,lenthe:jcp1994,lenthe:jcp1996} Hamiltonians. Various flavors of the ZORA Hamiltonian have been implemented in DIRAC,\cite{visscher:jcp2000} but with limited applicability. 
DIRAC features the very first implementation of a 2-component relativistic Hamiltonian that allows the \textit{exact} reproduction of the positive-energy spectrum of the parent 4-component
Hamiltonian.\cite{jensen:rehe2005,ilias:cpl2005} This implementation, presented as the BSS Hamiltonian by Jensen and Ilia\v{s}, referring to previous work by  Barysz, Sadlej and Snijders,\cite{barysz:ijqc1997}
carried out a free-particle Foldy-Wouthuysen transformation\cite{foldy:wouthuysen:pr1950} on the Dirac Hamiltonian, followed by an exact decoupling of positive- and negative-energy solutions.
This two-step approach allows the construction of finite-order 2-component relativistic Hamiltonians such as the first- and second-order Douglas--Kroll--Hess Hamiltonians,
but is unnecessary for exact decoupling. The code was therefore superseded by a simple one-step approach, reported as the Infinite-Order Two-Component Hamiltonian (IOTC) by Saue and Ilia\v{s}.\cite{ilias:jcp2007} Due to the equivalence with the exact quasirelativistic (XQR) Hamiltonian reported by Kutzelnigg and Liu\cite{kutzelnigg:jcp2005} it was later agreed\cite{x2c:2007}
to name such Hamiltonians eXact 2-Component (X2C) Hamiltonians. The X2C decoupling transformation is available in matrix form and is used to transform any one-electron integral to 2-component form,
hence avoiding picture change errors. For the two-electron integrals, DIRAC employs the uncorrected two-electron operator supplemented with the Atomic Mean-Field approach for including two-electron spin-orbit interaction.\cite{Hess_CPL1996,prog:amfi}

DIRAC features, in addition, one-component scalar relativistic effective core potentials (AREP) as well as two-component spin-orbit relativistic effective core potentials (SOREP).\cite{Park2012}

For wave function-based correlation methods, the electronic Hamiltonian is conveniently written in normal-ordered form
\begin{equation}
  \hat{H}=E^{HF}+\sum_{pq}F^p_q\{a^{\dagger}_pa^{\phantom{\dagger}}_q\}+\frac{1}{4}\sum_{pqrs}V^{pr}_{qs}\{a^{\dagger}_pa^{\dagger}_ra^{\phantom{\dagger}}_sa^{\phantom{\dagger}}_q\},
\end{equation}
where $E^{HF}$ is the Hartree--Fock energy, $F^p_q$ elements of the Fock matrix and curly brackets refer to normal ordering with respect to the Fermi vacuum, given by the HF determinant.
For such calculations DIRAC features the X2C molecular mean-field approach\cite{sikkema:jcp2009}: After a 4-component relativistic HF calculation, the X2C exact decoupling is carried out on
the Fock matrix, rather than the Dirac Hamiltonian matrix, whereas the two-electron operator is left untransformed. In combination with the usual approximation of neglecting core electron correlation, this limits the effect of picture change errors to valence-valence electron interactions only: core-core and core-valence electron interactions are treated with the same accuracy as in the 4C approach.

\subsection{Electronic structure models}
\subsubsection{Self-consistent field (SCF) calculations}
At the core of DIRAC is an SCF module allowing both Hartree--Fock (HF)\cite{Saue:MP1997} and Kohn--Sham (KS)\cite{saue:jcc2002} calculations. These calculations are Kramers-restricted and use a symmetry scheme based on quaternion algebra which automatically provides maximum point group and time-reversal symmetry reduction of the computational effort.\cite{Saue:Jensen:JCP1999} In nonrelativistic quantum chemistry codes, spin-restricted open-shell SCF calculations employ Configuration State Functions (CSFs) $|S,M_S\rangle$ of well-defined spin symmetry. However, in the relativistic domain spin symmetry is lost, and so the use of CSFs would require linear combinations of Slater determinants adapted to combined spin and spatial symmetry, which is a challenge for a general molecular code. We have therefore instead opted for the average-of-configuration Hartree-Fock method\cite{Thyssen2004} for open-shell systems. Individual electronic states may subsequently be resolved by a Complete Open-Shell CI calculation.\cite{Visser1992} Open-shell Kohn--Sham calculations use fractional occupation.

SCF calculations are based on the traditional iterative Roothaan--Hall diagonalization method with direct-inversion-in-the-iterative-subspace (DIIS) convergence acceleration. By default, the start guess is provided by a sum of atomic LDA potentials, which have been prepared using the GRASP atomic code\cite{prog:grasp} and are fitted to an analytical expression.\cite{lehtola:2020} Other options include i) bare nucleus potentials corrected with screening factors based on Slater's rules,\cite{Slater_PhysRev1930} ii) atomic start based on densities from atomic SCF runs for the individual centers\cite{vanLenthe:2006} and iii) an extended H{\"u}ckel start based on atomic fragments.\cite{Saue_adhoc} In each SCF iteration orbitals are by default ordered according to energy, and orbital classes are assigned by simple counting in the order: (secondary) negative-energy orbitals, inactive (fully occupied) orbitals, active (if any) orbitals and virtual orbitals. The implicit assumption of relative ordering of orbital energies according to orbital classes may cause convergence problems, for instance for \textit{f}-elements where closed-shell \textit{ns}-orbitals typically have higher energies than open-shell \textit{(n-2)f} ones. Such convergence problems may be avoided by reordering of orbitals combined with overlap selection, pioneered by Paul Bagus in the 1970s\cite{Bagus_JCP1971} and nowadays marketed as the Maximum Overlap Method.\cite{Gilbert_JPCA2008} Overlap selection also  provides robust convergence to core-ionized/excited states.\cite{saue:uraniumX2015} Negative-energy orbitals are treated as an orthogonal complement, corresponding to the implicit use of a projection operator.\cite{Almoukhalalati2016} 

An extensive selection of exchange-correlation energy functionals, as well as their derivatives to high order, needed for property calculations, are available for Kohn--Sham calculations. These XC functional derivatives are either provided by a module written by Sa{\l}ek\cite{Salek:2007} using symbolic differentiation, or by the \texttt{XCFun} library written by Ekstr{\"o}m~\cite{ekstrom-2010,xcfun} using the automatic differentiation technique. By default, Kohn--Sham calculations employ Becke partitioning\cite{Becke:1988} of the molecular volume into overlapping atomic volumes, where the numerical integration within each atomic volume is carried out using the basis-set adaptive radial grid proposed by Lindh, Malmqvist and Gagliardi\cite{Lindh2001} combined with angular Lebedev quadrature. XC contributions to energy derivatives or Fock matrix elements are evaluated for a batch of points at a time which allows us to screen entire batches based on values of basis functions at these grid points and enables us to express one summation loop of the numerical integration as a matrix-matrix multiplication step.

\subsubsection{Correlation methods}\label{subsec:corr-meth}

\paragraph{4-index transformations}
While the AO-to-MO index transformations are subordinate to the correlation approaches described below, some features are worth describing in a separate section. Irrespective of the Hamiltonian that is used in the orbital generation step, the approach always assumes that a large atomic orbital basis set is condensed to a much smaller molecular orbital basis. The result is a second-quantized, no-pair Hamiltonian in molecular orbital basis that is identical in structure to the second-quantized Hamiltonian encountered in nonrelativistic methods (see Eq.~\eqref{eq:secham}). The main difference is the fact that the defining matrix elements in this Hamiltonian are in general complex due to the inclusion of spin-orbit coupling in the orbital generation step. As a consequence, integrals will not exhibit the usual 8-fold permutation symmetry familiar from nonrelativistic integrals. This is even the case when higher point group symmetry is used to render the integrals real, as they may be a product of two imaginary transition densities. Only for spin-free calculations is it possible to choose phase factors for the spinors in such a way that 8-fold permutational symmetry is recovered. For ease of interfacing with nonrelativistic correlation implementations, such phase factors are inserted in the final stage of the transformation when running in spin-free mode. The primary interface files that are generated contain the (effective) one-body operator plus additional symmetry and dimensionality information needed to set up the Hamiltonian. The numerous 2-body matrix elements are stored in separate files that can be distributed over multiple locations in a compute cluster environment with delocalized disk storage. The program is complemented by a utility program that can convert the MO integrals to formats used by other major quantum chemistry programs, such as the FCIDUMP format,\cite{Knowles_CPC1989} thus facilitating the interfacing\cite{Nataraj2010} of DIRAC to other electron correlation implementations, such as MRCC\cite{mrcc}, or even to quantum computers. With respect to the latter, a 4-component relativistic quantum algorithm was reported in Ref.\citenum{veis2012relativistic}. More recently, DIRAC has been interfaced to the electronic structure package OpenFermion through a Python interface,\cite{senjean_openfermiondirac} thus allowing the calculation of energies and energy derivatives on a quantum computer\cite{obrien2019calculating} using either the full Dirac--Coulomb Hamiltonian or the L{\'e}vy--Leblond Hamiltonian.

The implementation has been revised several times over the years to account for changes in computer hardware architectures. The current default algorithm for the most demanding transformation of the 2-electron integrals uses an MPI type of parallelization in which half-transformed integrals are generated from recomputed AO integrals. If total disk space is an issue it is also possible to employ a scheme in which only a subset of half-transformed integrals is stored at a given time. The generation of one-body integrals is less demanding and carried out by calling the Fock matrix build routine from the SCF part of the program, with a modified density matrix that includes only contributions from the orbitals that are to be frozen in the correlation treatment. In the generation of these integrals it is possible to account for the Gaunt correction to the Coulomb interaction, thereby making a mean-field treatment of this contribution possible.\cite{sikkema:jcp2009} Explicit transformation of additional integrals over operators needed for the evaluation of molecular properties, or the inclusion of a finite strength perturbation in only the electron correlation calculation, is also possible and handled by a separate submodule. 

The lowest-level correlation method is second-order M{\o}ller-Plesset perturbation theory (MP2) and an early integral-direct, closed-shell implementation was realized by Laerdahl\cite{Laerdahl1997} in 1997.  This
implementation focuses on efficient, parallel calculation of the MP2 energy for closed shell 
systems. A more general implementation that also allows for calculation of the relaxed MP2 density matrix was realized later by van Stralen\cite{vanStralen:2005p1408} as part
of the coupled cluster implementation discussed below. Both implementations use the conventional MP2 approach; a more efficient Cholesky-decomposed density matrix implementation was developed by Helmich-Paris\cite{HELMICHPARIS201938}. In this approach, the quaternion formalism, which has been developed in earlier works,\cite{HELMICHPARIS2016} was used to reduce the number of operations. A production implementation along these lines is planned for the 2020 release.

\paragraph{Configuration Interaction}
The first implementation (\textbf{DIRRCI} module) of Restricted Active Space Configuration Interaction was taken from the MOLFDIR program\cite{prog:molfdir,MOLFDIR-source} and is briefly described in sections 3.4 and 4.10 of Ref.~\citenum{prog:molfdir}, with more details on the calculation of CI coupling coefficients given in Chapter 6.5 of Ref.~\citenum{visscher:TCC2002}. This module is mostly kept for reference purposes as a more general implementation of configuration interaction in DIRAC was introduced later by Fleig and coworkers.\cite{fleig_gasci2} A unique feature that makes the DIRRCI module still of some interest is the handling of Abelian point group symmetry. The implementation is capable of handling every possible Abelian group as long as the respective multiplication table is provided. This feature allows for treatment of linear symmetry (a feature lacking in the MOLFDIR program) by merely changing the dimensions of the arrays that hold the symmetry information. While the DIRRCI code is
no longer actively developed, some later extension beyond the MOLFDIR capabilities have been implemented, like the (approximate) evaluation of correlated first-order properties using the unrelaxed CI density matrix by Nayak.\cite{Nayak2006,Nayak2007,Nayak2009} This implementation allows for the calculation of expectation values of one-electron property operators over the CI wave functions.

The more recent \textbf{KR-CI} module is a string-based Hamiltonian-direct configuration interaction (CI) program that uses Dirac Kramers pairs from either a closed- or an open-shell calculation in a relativistic two- and four-component formalism exploiting double point-group
symmetry. KR-CI is parallelized using MPI in a scalable way where the CI vectors are distributed over the nodes, thus enabling use
of the aggregate memory on common computing clusters .\cite{Knecht2009,Knecht2010a} There are two choices for the CI kernel: LUCIAREL or GASCIP.

The LUCIAREL kernel \cite{fleig_gasci2,fleig_gasmcscf} is a relativistic generalization
of the earlier LUCIA code by Olsen.\cite{olsen_billion}
It is capable of doing efficient CI computations at arbitrary excitation level, FCI, SDCI, RASCI, and MRCI, all of which are subsets of the Generalized Active Space (GAS) CI. The
GAS concept \cite{olsen_cc} is a central and very flexible feature (described in
greater detail in Ref. \citenum{hubert_LRCC_ScH_JCP2013}) of the program that
can be applied effectively to describe various physical effects in atomic matter.
Apart from routine applications to valence electron correlation 
\cite{gomes_meth_triiodide,rota_fleig_etal_zfs} it has been used in other modern applications to efficiently describe core-valence electron correlation \cite{fleig:PRA2016} and also core-core correlation.\cite{Fleig_PRA2019} 
These uses can be combined with excited-state calculations, even when greater 
numbers of excited states with varying
occupation types are required.\cite{FleigJung_JHEP2018} In the latter case typical CI
expansion lengths are on the order of up to $10^8$ terms, whereas parallel 
single-state calculations have been carried out with up to $10^{10}$ expansion terms.\cite{Denis-Fleig_ThO_JCP2016}
If desired, KR-CI computes 1-particle densities from optimized CI wave functions from which the natural orbital occupations are deduced.

Based on the original complete open-shell implementation (GOSCIP) of MOLFDIR that is used to obtain state energies after an average-of-configuration Hartree-Fock calculation a  more general and efficient \textbf{GASCIP} (Generalized Active Space CI Program) module was originally written by Thyssen and Jensen for KR-MCSCF and later parallelized by Jensen. This specialized CI implementation is primarily used for KR-MCSCF and ESR calculations,\cite{Vad:2013} but it is also available for KR-CI calculations. Another separate implementation is the spin-free version of LUCIAREL, \textbf{LUCITA}, which fully exploits spin and boson symmetry.
LUCITA will consequently be faster for spin-free CI calculations than KR-CI using the LUCIAREL kernel. LUCITA has also been parallelized with MPI,\cite{Knecht2008} in the same way as KR-CI.

\paragraph{Multiconfigurational SCF}\label{subsec:krmcscf}

The original KR-MCSCF implementation was written by Thyssen, Fleig and Jensen\cite{Thyssen2008}
and follows closely the theory of Ref.\ \citenum{Jensen1996}. Within a given symmetry sector, it allows for a state-specific optimization by taking advantage of a Newton-step-based genuine second-order optimization algorithm.\cite{Jensen1996}    
The KR-MCSCF module was later parallelized\cite{Knecht2009,Knecht2010a} using MPI by Knecht, Jensen and Fleig by means of a parallelization of the indiviual CI-based tasks encountered in an MCSCF optimization: (i) generation of start vector, (ii) sigma-vector calculation and (iii) evaluation of one- and two-particle reduced density matrices (RDMs).  Moreover, its extension to an efficient treatment of linear symmetry (see Section \ref{subsec:symmetry-considerations}) -- both in the KR-CI part and the restriction of orbital-rotation parameters --  was a central element that allowed a comprehensive study\cite{knecht_u2} of the electronic structure as well as of the chemical bond in the ground- and low-lying excited states of U$_2$  based on a simultaneous, variational account of both (static) electron correlation and spin-orbit coupling.

\paragraph{Coupled Cluster\label{relccsd}}

The \textbf{RELCCSD} coupled cluster module\cite{Visscher:1996p105} is also derived from the MOLFDIR implementation, but in contrast to the DIRRCI module is still under active development. The implementation uses the same philosophy as DIRRCI in demanding only a point group multiplication table to handle Abelian symmetry. Point group symmetry beyond Abelian symmetry is used to render the defining integrals of the second quantization Hamiltonian real, using the scheme first outlined in Ref. \citenum{Visscher:1996p778}, but adjusted to work with the quaternion algebra used elsewhere in DIRAC. The implemented algorithms work in the same way for real and complex algebra, with all time-consuming operations performed by BLAS\cite{Lawson:1979dr} calls that are made via a set of wrapper routines that point to the (double precision) real or complex version, depending on the value of a global module parameter. In this way the need for maintenance of separate code for real or complex arithmetic is strongly reduced. Due to the dependence on BLAS operations, shared memory parallelization is easily achieved by linking in a multithreaded BLAS library. Parallelization via MPI can be achieved in addition, as described in Ref. \citenum{Pernpointner:2003p1085}.

For the description of electronic ground states that can be qualitatively well described by a single determinant, the standard CCSD(T) model is usually the optimal choice in terms of performance, with the code taking the trial CCSD amplitudes from an MP2 calculation. As the RELCCSD implementation does not assume time-reversal symmetry, it is possible to treat open shell cases as well. This is straightforward at the CCSD level of theory as the precise open-shell SCF approach used to generate the orbitals is then relatively unimportant and the implementation does not assume a diagonal reference Fock matrix. For the implemented perturbative triple corrections\cite{Raghavachari:1989,Deegan:1994vd} a larger dependence on the starting orbitals is observed, although performance is usually satisfactory for simple open shell cases with only one unpaired electron. For more complicated cases it is often better to use the Fock space coupled cluster model (FSCC), in which multireference cases can also be handled. 

The relativistic use of the FSCC method \cite{Lindgren:1987in} has been pioneered by Eliav and Kaldor,\cite{Eliav:1994hx} and in the DIRAC implementation\cite{Visscher:2001p87} one is able to investigate electronic states which can be accessed by single or double electron attachment or detachment, as well as singly excited states, from a starting closed-shell reference determinant---that is, states with up to two unpaired electrons. Most calculations done with this method nowadays use their intermediate Hamiltonian\cite{Landau:JCP2000,Eliav:2005extrapolated} (IH) schemes that remove problems with intruder states to a large extent\cite{Infante:2006p1405,actinide-Infante-JCP2007-127-124308,actinide-Real-JPC2009-113-12504,actinide-Tecmer-PCCP2011-13-6249}. The IH approach has been recently accomplished by the very efficient Pad{\'e} extrapolation method\cite{Zaitsevskii:2018pade}. As IH schemes often use large active spaces, and the original Fock space implementation\cite{Visscher:2001p87} was designed with small numbers of occupied orbitals in mind, these calculations can become rather memory-intensive.
A less demanding scheme based on the equation-of-motion coupled cluster approach has been 
implemented\cite{Shee:2018bg} for the treatment of electron attachment (EOM-EA), ionization (EOM-IP) and excitation energies (EOM-EE). EOM-IP and EOM-EE can also be used to obtain core ionization and excitation energies via the core-valence separation approach.\cite{halbert2020cvs}  This complements the Fock space functionality for treating electronically excited states, especially for species which can be represented by a closed-shell ground-state configuration.\cite{actinide-Kervazo-IC2019}

\paragraph{Range-separated density functional theory} This method allows grafting of  wave function-based correlation methods onto density functional theory without double counting of electron correlation. We have explored this approach combining short-range DFT with long-range MP2\cite{Kullie_CP2012} and CC.\cite{Shee_PCCP2015}

\paragraph{Density Matrix Renormalization Group.} If requested, KR-CI provides a one- and two-electron integral file in FCIDUMP format\cite{Knowles_CPC1989} which allows, for the active space specified in the KR-CI input, relativistic density matrix renormalization group (DMRG) calculations \cite{Knecht_2014,Battaglia_2018} with the relativistic branch of the DMRG program \textsc{QCMaquis}.\cite{Keller_2015,Knecht_2016} If desired, the DMRG program computes the one-particle reduced density matrix for the optimized wave function in the MO basis and writes it to a text file which can be fed back into DIRAC. This feature makes it possible to calculate  (static) first-order one-electron properties in the same way as described below  for SCF calculations. Moreover, this functionality also opens up further possibilities to analyze the resulting wave function and to calculate  properties in real-space as it was shown in Ref.~\citenum{Battaglia_2018}\ for a Dy(III) complex. For the latter functionality, the visualization module (see Section \ref{visualization}) was extended with an interface to wave functions optimized within the KR-CI/KRMCSCF framework of Dirac.    

\subsection{Molecular properties}
A common framework for the calculation of molecular properties is  response theory. A less economical, but computationally easier, approach is to use a finite-field approach. Implementations for both strategies are available in DIRAC.

The property module of DIRAC has been written in a very general manner, allowing the user to define 4-component property operators, benefiting from the extensive integral menu of the HERMIT module.\cite{prog:hermit} For convenience, a number of properties are predefined, as shown in Table \ref{tab:preprop}.

\begin{table}[ht]
\caption{\label{tab:preprop}Predefined molecular properties in DIRAC}
    \centering
\scriptsize
\begin{tabular}{|l|l|c|c|c|c|}
\hline\hline
Keyword&\textit{Electric properties} & EV & LR & QR & Refs.\tabularnewline
\hline
\texttt{.DIPOLE} & Electric dipole moment & x &  &  & \tabularnewline
\texttt{.QUADRU} & Traceless electric quadrupole moment & x &  &  & \tabularnewline
\texttt{.EFG} & Electric field gradients at nuclear positions & x &  &  & \citenum{Visscher_JCP1998}\tabularnewline
\texttt{.NQCC} & Nuclear quadrupole coupling constants & x &  &  & \citenum{Visscher_JCP1998}\tabularnewline
\texttt{.POLARI} & Electronic dipole polarizability tensor &  & x &  & \citenum{Saue:Jensen:JCP2003,Salek2005}\tabularnewline
\texttt{.FIRST } & Electronic dipole first-order hyperpolarizability tensor &  &  & x & \citenum{Norman_JCP2004}\tabularnewline
\texttt{.TWO-PH} & Two-photon absorption cross sections &  &  & x & \citenum{Henriksson:2005}\tabularnewline
\hline
&\textit{Magnetic properties}\tabularnewline
\hline
\texttt{.NMR} & Nuclear magnetic shieldings and indirect spin-spin couplings  &  & x &  & \citenum{Visscher_jcc1999,Ilias:JCP2009}\tabularnewline
\texttt{.SHIELD} & Nuclear magnetic shieldings &  & x &  & \citenum{Visscher_jcc1999,Ilias:JCP2009}\tabularnewline
\texttt{.SPIN-S} & Indirect spin-spin couplings &  & x &  & \citenum{Visscher_jcc1999}\tabularnewline
\texttt{.MAGNET} & (static) Magnetizablity tensor &  & x &  & \citenum{Ilias:MP2013,Olejniczak:dftmagn}\tabularnewline
\texttt{.ROTG} & Rotational g-tensor (DIRAC20) &  & x &  & \citenum{Aucar:JCP2014}\tabularnewline
\hline
&\textit{Mixed electric and magnetic properties}\tabularnewline
\hline
\texttt{.OPTROT} & Optical rotation &  & x &  &\citenum{Creuzberg_ecd} \tabularnewline
\texttt{.VERDET} & Verdet constants &  &  & x & \citenum{Ekstrom2005}\tabularnewline
\hline
&\textit{Other predefined properties}\tabularnewline
\hline
\texttt{.MOLGRD} & Molecular gradient &  &  &  &\citenum{Thyssen_molgrd} \tabularnewline
\texttt{.PVC} & Parity-violating energy (nuclear spin-independent part) & x &  &  & \citenum{Laerdahl:PRA1999,Bast2011}\tabularnewline
\texttt{.PVCNMR} & Parity-violating contribution to the NMR shielding tensor &  & x &  & \citenum{Bast:JCP2006}\tabularnewline
\texttt{.RHONUC} & Electronic density at the nuclear positions (contact density) & x &  &  & \citenum{Knecht2011}\tabularnewline
\texttt{.EFFDEN} & Effective electronic density associated with nuclei (Mössbauer) & x &  &  & \citenum{Knecht2011}\tabularnewline
\texttt{.SPIN-R} & Nuclear spin-rotation constants & x & x &  & \citenum{Aucar:JCP2012}\tabularnewline
\hline
\end{tabular}
\end{table}

\subsubsection{SCF calculations}
At the closed-shell SCF level, DIRAC allows the calculation of molecular properties corresponding to expectation values as well as linear\cite{Saue:Jensen:JCP2003,Salek2005} and quadratic\cite{henriksson:2008,bast:JCP2009} response functions. 
In addition, first- and second-order residues of the quadratic response function have been programmed, allowing the calculation of two-photon absorption cross sections\cite{Henriksson:2005} and first-order properties of electronically excited states\cite{Tellgren_JCP2007}, respectively. 

Linear response functions have been extended to complex response through the introduction of a common damping term that removes divergences at resonances.\cite{Villaume_JCP2010} This allows not only probing of second-order properties in the vicinity of resonances, but also simulation of absorption spectra within a selected window of frequencies. In addition, complex response allows the calculation of properties at formally imaginary frequencies, such as C$_6$ dispersion coefficients.\cite{Sulzer_MP2012}

Excitation energies are available through time-dependent HF and DFT.\cite{Bast_ijqc2009} Restrictions may be imposed on active occupied (and virtual) orbitals, hence allowing restricted excitation window (REW) calculations\cite{Stener_CPL2003,saue:uraniumX2015} of X-ray absorption spectra. Another method available for core-excitation processes in molecules is the static-exchange approximation (STEX).\cite{Ekstrom_PhysRevA2006} Transition moments may be calculated with user-specified property operators. From DIRAC20 and onwards three schemes\cite{List:JCP2020} to go beyond the electric-dipole approximation in the calculation of oscillator strengths will be available in DIRAC. The first is based on the full semi-classical light-matter interaction operator, and the two others on a truncated interaction within the Coulomb gauge (velocity representation) and multipolar gauge (length representation). The truncated schemes can be calculated to \textit{arbitrary} order in the wave vector. All schemes allow rotational averaging.

NMR shieldings as well as magnetizabilities may be calculated using London orbitals and simple magnetic balance.\cite{Olejniczak:JCP2012} For KS calculations non-collinear spin magnetization has been implemented\cite{bast:JCP2009} and all required derivatives of exchange-correlation functionals are provided.

\subsubsection{Correlation modules}\label{sec:corrwav}

\paragraph{Electronic ground state properties}
The implementations for obtaining density matrices for molecular properties are still under development. The currently available functionality is to obtain the unrelaxed one-particle density matrix for the single reference CCSD model.\cite{Shee:2016fv} For the MP2 model, for which orbital relaxation effects are more important, the relaxed density matrix can be obtained.\cite{vanStralen:2005p1408} After back-transforming to the AO basis, molecular properties can be obtained in the same way as for SCF calculations. Alternatively one may also obtain matrix elements of property operators in the MO basis and compute
expectation values of CI wave functions and/or include 
property operators as a finite-strength perturbation in the
CI or CC wave function determination. This allows determination of properties that break Kramers symmetry and has for example been used for assessing the effect of an electric dipole moment of the electron ($e$EDM) in molecular systems.\cite{Denis:2019bx}

\paragraph{Excited state properties}
For properties that depend also on the excited
state density matrix, such as transition probabilities, only limited functionality is available in RELCCSD. Transition intensities based on an approximate CI expression and the
dipole approximation for the transition moments have been
implemented for the Fock space coupled cluster model. 
Under development and planned to be available in the 2020 DIRAC release, is an extension to non-diagonal form of the finite field approach in Fock Space CC. This approach allows accurate calculations of the dipole moments of electron transitions in heavy atomic and molecular systems\cite{Zaitsevskii:2018electronic}.  
For KR-CI the range of properties is larger\cite{Knecht2009}. 
These include molecule-frame static electric dipole moments,\cite{Knecht2009,Denis2015} E1 transition matrix elements \cite{Knecht2009,PhysRevA.95.022504} and magnetic hyperfine interaction constants
\cite{Fleig2014} in electronic ground and excited states. Moreover, parity- (P) and
time-reversal (T) violating properties are implemented as expectation values over 
atomic or molecular KR-CI ground- and excited-state wavefunctions, in particular the electron electric
dipole moment interaction,\cite{Fleig2013} the P,T-odd scalar-pseudoscalar nucleon-electron interaction \cite{Denis2015} and the nuclear magnetic quadrupole moment electron magnetic-field interaction.\cite{fleig:PRA2016}


\paragraph{Electron propagator}
The Algebraic Diagrammatic Construction (ADC) is an efficient, size-extensive post-Hartree-Fock method, which can be used to obtain molecular properties.
With DIRAC the calculation of single\cite{Pernpointner04_2} and double\cite{Pernpointner10_1} ionization as well as electronic excitation\cite{Pernpointner14,Pernpointner18} spectra using the RELADC and POLPRP modules are possible.
Decay widths of electronic decay processes can be obtained
by the FanoADC-Stieltjes method.\cite{Fasshauer15_1}
The ionization spectra can be obtained at the level of ADC(2), ADC(2x) as well as ADC(3) plus constant diagrams, while the electronic excitation spectra are available at ADC(2) and ADC(2x) levels of accuracy.
Technically, the ADC implementation and the RELCCSD code share much of their infrastructure.

\paragraph{Quasi-degenerate perturbation theory using configuration interaction}
A module is under development for description of properties of quasi-degenerate states as encountered in open-shell molecules.
The focus has so far been on calculation of ESR/EPR g-tensors,\cite{Vad:2013} hyperfine couplings and zero-field splitting.
It is based on the flexible GASCIP configuration interaction module.

\subsection{Environments}\label{subsec:embedding-mod}

A great deal of information can be extracted from gas-phase electronic structure calculations, even for systems that are studied experimentally in solution or other condensed phases. Nevertheless, the environment can strongly modulate the properties of a system, such as formation/reaction energies and response properties (e.g.~electronic or vibrational spectra).
It can therefore be important to take into account the influence of the environment on the systems of interest.
A straightforward way to include the environment is by performing calculations on large molecular systems or aggregates, but already for HF and DFT calculations that quickly becomes unwieldy, and for correlated electronic structure calculations (section~\ref{sec:corrwav}) this strategy is largely unfeasible.

\begin{figure}[h]
  \centering
  \includegraphics[width=.5\textwidth]{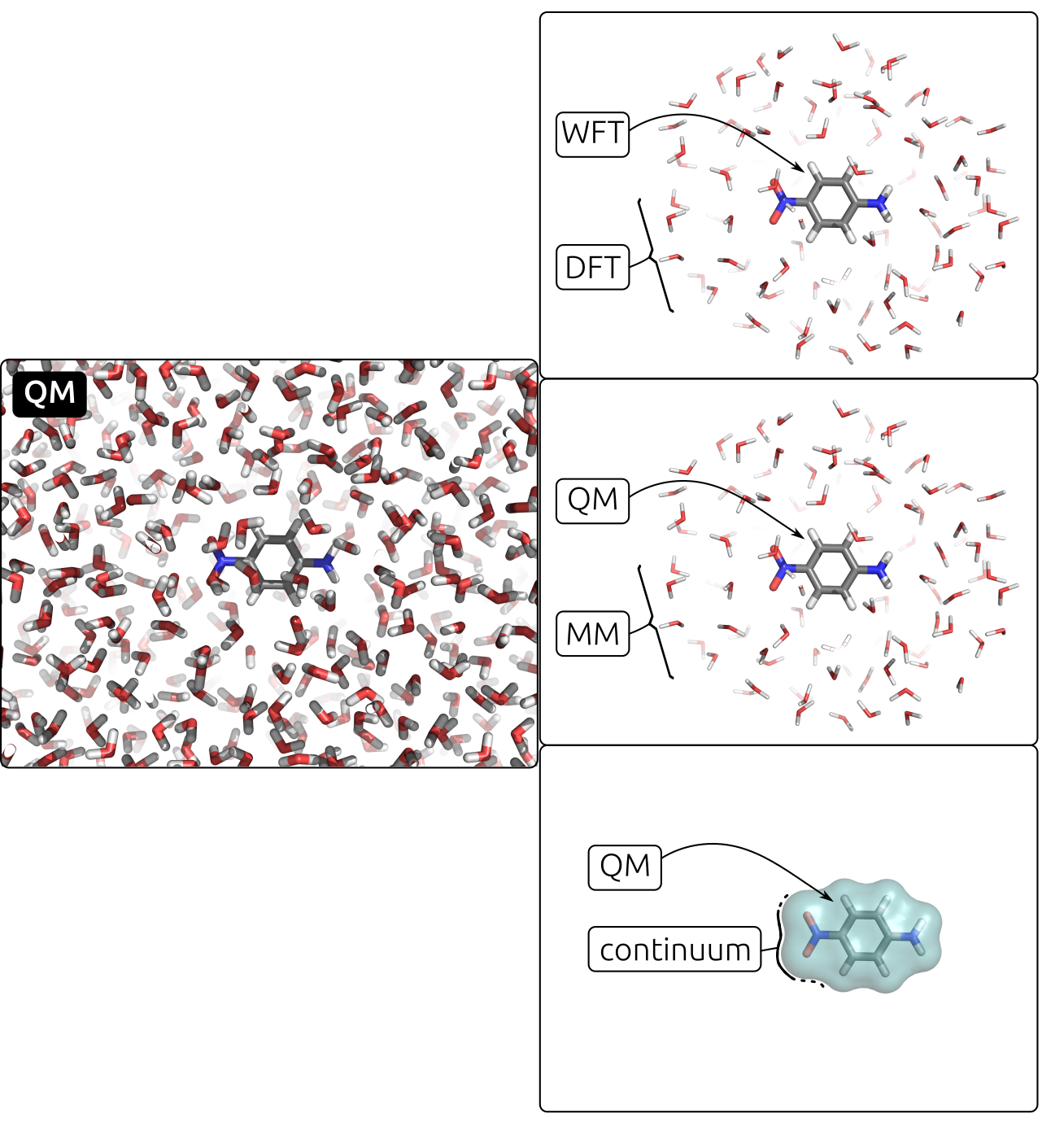}
  \caption{Pictorial depiction of the transition from \textbf{quantum
mechanical} to one of the \textbf{multiscale} models available in DIRAC for the
aqueous solvation of \emph{para}-nitroaniline. Leftmost panel: a \textbf{fully
quantum} mechanical cluster model. Upper central panel: a \textbf{frozen density
embedding} (explicit) model. Middle central panel: a \textbf{quantum/classical
discrete} (explicit) model. Lower central panel: a \textbf{quantum/classical
continuum} (implicit) model.}\label{fig:multiscale}
\end{figure}

The alternative to such calculations is the use of embedding
approaches,\cite{SeveroPereiraGomes2012} in which the environment is treated in
a more approximate fashion with respect to the subsystem of
interest,  see Figure~\ref{fig:multiscale}. Apart from the simplest possible embedding scheme, using \emph{fixed} point
charges, DIRAC offers three different classes of increasingly sophisticated
embedding approaches: the implicit solvation polarizable continuum model (PCM),
the atomistic polarizable
embedding (PE) model, and the frozen density embedding (FDE) model; the latter also referred to as
subsystem DFT.
In the first two, the environment is treated classically, whereas
for the latter all fragments are treated quantum mechanically, though generally
at different levels of theory.

\subsubsection{Polarizable continuum model (PCM)\label{pcm}}

The PCM model is a quantum/classical polarizable model for the approximate
inclusion of solvent effects into quantum mechanical calculations.\cite{Tomasi2005-vm}
It is a focused, implicit solvation model: the environment (usually a solvent) is
replaced by a dielectric with permittivity $\varepsilon$ and the mutual
interaction between the quantum mechanical and classical regions is described by
the electrostatic polarization. The model cannot describe specific, weak
interactions between subsystems, such as hydrogen bonding, but can give a first
qualitative estimate of solvation effects on many molecular properties.

The quantum mechanical region is delimited by a \emph{cavity}, a
region of space usually constructed as a set of interlocking spheres and hosting
the QM fragment, see lower central panel in Figure~\ref{fig:multiscale}.
Inside the cavity, the permittivity is that of vacuum ($\varepsilon=1$), while
outside the permittivity assumes the value appropriate for the environment being
modeled. For example, in the case of water the experimental value $\varepsilon=78.39$ would be used.
The electrostatic polarization is represented as an apparent surface charge
(ASC), which is the solution to the integral formulation of
the Poisson equation. The coupling with the QM code at the SCF level of theory is achieved by augmenting
the usual Fock operator with an ASC-dependent environment polarization operator.
This results in minimal, localized changes to the SCF cycle.

The implementation in DIRAC is based on the~\textsc{PCMSolver}
library,\cite{diremigio2019, Di_Remigio2015-dt} which
provides a well-defined interface to a stand-alone computational backend. The
PCM method is available for mean-field (Hartree--Fock and Kohn--Sham DFT)
wavefunctions. Additionally, static electric linear response properties can also
be computed including the effect of the solvent \emph{via} the PCM.

\subsubsection{Polarizable Embedding (PE)\label{pe}}
The PE model is a fragment-based quantum--classical embedding approach for including environment effects in calculations of spectroscopic properties of large and complex molecular systems.\cite{Olsen2010,Olsen2011,Steinmann2018}
The effects from the classical environment on the quantum subsystem are included effectively through an embedding potential that is parameterized based on \textit{ab initio} calculations.
The molecular environment is thus subdivided into small, computationally manageable fragments from which multi-center multipoles and multi-center dipole--dipole polarizabilities are computed.
The multipoles and polarizabilities model the permanent and induced charge distributions of the fragments in the environment, respectively.
For solvent environments, a fragment typically consists of an individual solvent molecule, while for large molecules, such as proteins, a fragmentation approach based on overlapping fragments is used.
The resulting embedding potential is highly accurate\cite{Olsen2015} and introduces an explicitly polarizable environment that thus allows the environment to respond to external perturbations of the chromophore.\cite{List2016b}
The embedding-potential parameters can be conveniently produced using external tools such as the PyFraME package,\cite{pyframe} which automates the workflow leading from an initial structure to the final embedding potential.

The current implementation of the PE model in DIRAC can be used in combination with mean-field electronic-structure methods (i.e., HF and DFT) including electric linear response and transition properties where local-field effects, termed effective external field (EEF)\cite{List2016a,List2017} effects in the PE context, may be included.\cite{Hedegrd2017}
The model is implemented in the Polarizable Embedding library (PElib)\cite{pelib} that has been interfaced to DIRAC.\cite{Hedegrd2017}
The library itself is based on an AO density-matrix-driven formulation, which facilitates a loose-coupling modular implementation in host programs.
The effects from the environment are included by adding an effective one-electron embedding operator to the Fock operator of the embedded quantum subsystem.
The potential from the permanent charge distributions is modeled by the multipoles, which is a static contribution that is computed once and added to the one-electron operator at the beginning of a calculation.
The induced, or polarized, charge distributions are modeled by induced dipoles resulting from the electric fields exerted on the polarizabilities.
This introduces a dependence on the electronic density, through the electronic electric field, and the induced dipoles are therefore updated in each iteration of an SCF cycle or response calculation, similar to the procedure used in PCM.\cite{List2013}

\subsubsection{Frozen density embedding (FDE)\label{fde}}

The FDE approach is based on a reformulation of density functional theory whereby one can express the energy of a system in terms of subsystem energies and an interaction term (see Ref. \citenum{SeveroPereiraGomes2012} and references therein), which contains electrostatic, exchange-correlation and kinetic energy contributions, the latter two correcting for the non-additivity between these quantities calculated for the whole system and for the individual subsystems. As in other embedding approaches, we are generally interested in one subsystem, while all others constitute the environment. The electron density for the system of interest is determined by making the functional for the total energy stationary with respect to variations of the said density, with a constraint provided by the density of the environment. The interaction term thus yields a local embedding potential, $v^{\mathrm{emb}}(\mathbf{r})$, representing the interactions between the system and its environment.  

The FDE implementation in DIRAC is capable of calculating $v^{\mathrm{emb}}(\mathbf{r})$ during the SCF procedure (HF and DFT) using previously obtained densities and electrostatic potentials for the environment on a suitable DFT integration grid, as well as to export these quantities. One can also import a precalculated embedding potential, obtained with DIRAC or other codes,\cite{pyadf-2011} and include it in the molecular Hamiltonian as a one-body operator.\cite{actinide-gomes-pccp2008-10-5353} This allows for setting up iterative procedures to optimize the densities of both the system of interest and the environment via Freeze-Thaw cycles.\cite{env-Halbert-IJQC2020} 

At the end of the SCF step the imported or calculated $v^{\mathrm{emb}}(\mathbf{r})$ becomes part of the optimized Fock matrix, and is therefore directly included in all correlated treatments mentioned above~\cite{actinide-gomes-pccp2008-10-5353,halides-water-Bouchafra-PRL2019} as well as for TD-HF and TD-DFT. For the latter two, contributions arising from the second-order derivatives of interaction energy are also available for linear response properties for electric~\cite{env-Hofener-JCP2012-136-044104} and magnetic perturbations,\cite{Olejniczak2017} though the couplings in the electronic Hessian between excitations on different subsystems are not yet implemented. 
The interaction term for the non-additive kinetic energy contributions is calculated with one of the available approximate kinetic energy functionals that can be selected via input.

\subsection{Analysis and visualization}\label{visualization}
DIRAC features Mulliken population analysis.\cite{Mulliken:1955} However, this analysis should be used with caution due to its well-known basis-set dependence. An additional complication in the present case is that the analysis distributes density according to \textit{scalar} basis functions which is incompatible with 2- or 4-component atomic orbitals. We have therefore introduced \textbf{projection analysis}, similar in spirit to Mulliken analysis, but using precalculated atomic orbitals.\cite{Saue_CPL1996,Dubillard2006} The reference atomic orbitals may furthermore be polarized within the molecule using the Intrinsic Atomic Orbital algorithm.\cite{Knizia2013,Saue_iaos} The projection analysis furthermore allows the decomposition of expectation values at the SCF level into inter- and intra-atomic contributions, which for instance has elucidated the mechanisms of parity-violation in chiral molecules.\cite{Bast2011} It is also possible to localize molecular orbitals, which is favourable for bonding analysis.\cite{Dubillard2006}

The \textbf{visualization module} in DIRAC makes it possible to export densities and their derivatives, as well as other quantities (such as property densities obtained from response calculations) to third-party visualization software commonly used by the theoretical chemistry community such as Molden,\cite{molden00,molden17} as well as by less known analysis tools such as the Topology Toolkit (TTK),\cite{TTK} with which we can perform a wide range of topological analyses, including atoms-in-molecules (AIM)~\cite{analysis-Olejniczak-IJQC2019} with densities obtained with Hartree-Fock, DFT and CCSD wavefunctions. DIRAC can export such data in Gaussian cube file format, or over a custom grid.

DIRAC has been extensively used for the visualization of property densities, in particular magnetically induced currents.\cite{bast-cp-356-187-2009,sulzer-pccp-13-20682-2011} More recently, shielding densities have been investigated in order to gain insight into the performance of FDE for such NMR properties.\cite{Olejniczak2017,env-Halbert-IJQC2020}

\begin{figure}[h]
  \centering
  \begin{minipage}{0.5\textwidth}
    \centering
    \includegraphics[width=0.5\textwidth]{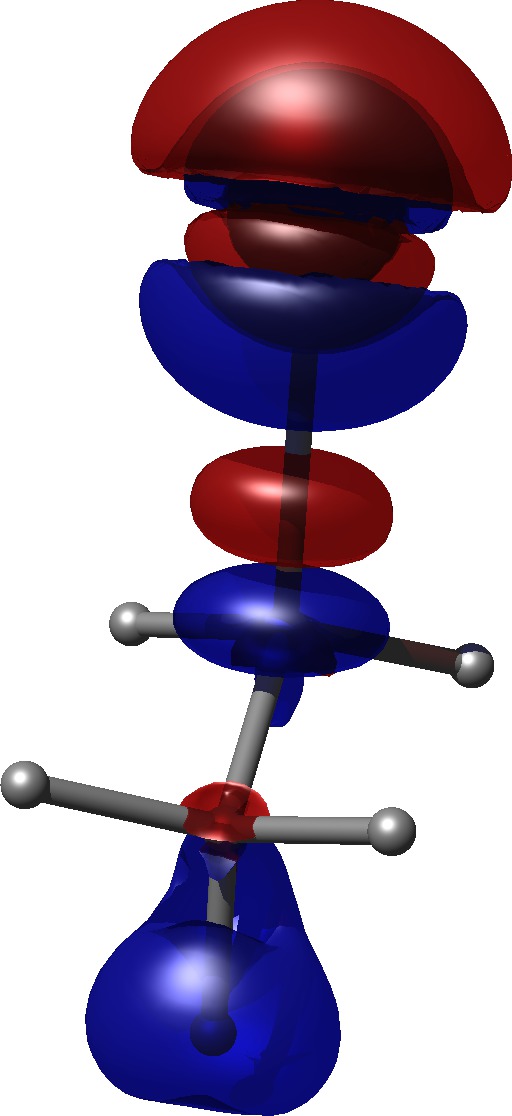}
  \end{minipage}%
  \begin{minipage}{0.5\textwidth}
    \centering
    \includegraphics[width=0.5\textwidth]{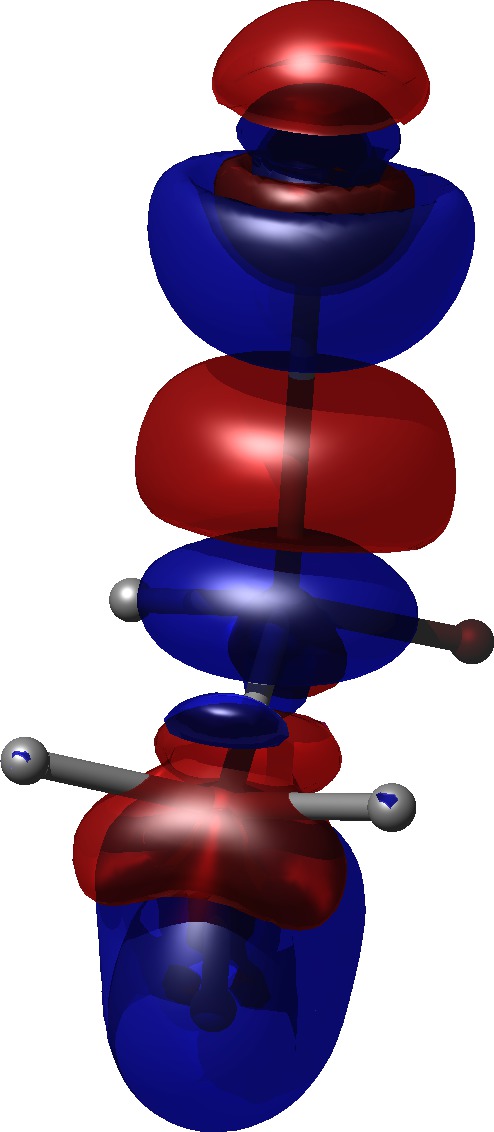}
  \end{minipage}
  \caption{Visualizing the analogy between the spin-orbit contribution to shieldings and indirect spin-spin coupling: Isosurface plot of the spin-spin coupling density $K(H_{\beta},I)$ in iodoethane (left panel; Fermi-contact + spin-dipole contributions), compared with the spin–orbit coupling contribution to the shielding density $\sigma(H_\beta)$ (right panel). The dihedral angle H--C--C--I is 180$^\circ$.}\label{fig:iodoethane}
\end{figure}

As an illustration of the visualization module, we start from the observation of Kaupp \textit{et al.}\cite{Kaupp_ChemEur1998} that the spin-orbit contribution to the shielding $\sigma(H_\beta)$ of the $\beta$-hydrogen of iodoethane follows closely the Karplus curve of the indirect spin-spin coupling constant $K(H_{\beta},I)$ as a function of the H--C--C--I dihedral angle. The DIRAC program makes it possible to isolate spin-free and spin-orbit contributions to magnetic properties;\cite{saue:aqc2005} this has allowed us to show that this connection is manifest at the level of the corresponding property densities (Figure~\ref{fig:iodoethane}).

\subsection{Programming details and installation}
\label{subsec:programming}

The source code consists mostly of Fortran 77 and Fortran 90 code, but some modules are written in C (exchange-correlation functional derivatives using symbolic differentiation, pre-Fortran-90 memory management) and C++ (exchange-correlation functional derivatives using automatic differentiation, polarizable continuum model). Python is used for the powerful code launcher \texttt{pam}, which has replaced the previous launcher written in Bash. 

The code base is under version control using Git and hosted on a GitLab repository server. The main development line as well as release branches are write-protected and all changes to these are automatically tested and undergo code review. For integration tests we use the Runtest library,\cite{runtest} and we run the test set both nightly as well as before each merge to the main code development branch.

Since 2011 the code is configured using CMake \cite{cmake} which was introduced to make the installation more portable and to make it easier to build and maintain a code base with different programming languages and an increasing number of externally maintained modules and libraries. The code is designed to run on a Unix-like operating system, but thanks to the platform universality of the employed Python, Git and CMake tools we have also been able to adapt the DIRAC code for the MS Windows operating system using the MinGW-GNU compilers suite. 
\subsection{Code documentation}
\label{subsec:documentation}

The code documentation (in HTML or PDF format) is generated from sources in reStructuredText format using Sphinx~\cite{sphinx} and served via the DIRAC program website.\cite{dirac-website} We track the documentation sources in the same Git repository as the source code. This way we are able to provide documentation pages for each separate program version, which improves the reproducibility of the code and also allows us to document unreleased functionality for future code versions. In addition to a keyword reference manual we share a broad spectrum of tutorials and annotated examples which provide an excellent starting point for users exploring a new code functionality or entering a new field.

\subsection{Distribution and user support}
\label{subsec:distribution}

The program is distributed in source code form under a custom open-use license. Traditionally we have distributed the code to the community upon request but starting with the DIRAC18 release~\cite{DIRAC18} we have switched to distributing the source code and collecting download metrics via the Zenodo service.\cite{zenodo} We plan to transition to an open source license (GNU Lesser General Public License) in the near future to encourage contributions and simplify derivative work based on the DIRAC package. User support is provided on a best-effort basis using Google Groups, with presently 360 subscribers. Deliberate efforts in community building are also reflected in the social media presence.\cite{dirac-twitter}

\section{Implementation Details}\label{sec:impdet}

\subsection{Basis functions}
In the nonrelativistic domain basis functions are modelled on atomic orbitals, but with adaptions facilitating integral evaluation. This has led to the dominant, but not exclusive, use of Cartesian or spherical Gaussian-type orbitals (GTOs). 4-component atomic orbitals may be expressed as
\begin{equation}\label{eq:relAO}
    \psi\left(\boldsymbol{r}\right)=
    \left[\begin{array}{c}\psi^L\\\psi^S\end{array}\right]
=\frac{1}{r}\left[\begin{array}{c}
P_{\kappa}(r)\xi_{\kappa,m_{j}}(\theta,\phi)\\
iQ_{\kappa}(r)\xi_{-\kappa,m_{j}}(\theta,\phi)
\end{array}\right],
\end{equation}
where $P_{\kappa}$ and $Q_{\kappa}$ are real scalar radial functions and $\xi_{\kappa,m_{j}}$ are 2-component complex angular functions. The first 4-component relativistic molecular calculations in the finite basis approximation met with failure because the coupling of the large $\psi^L$ and the small $\psi^S$ components through the Dirac equation was ignored. Since the exact coupling is formally energy-dependent, use of the nonrelativistic limit was made instead, leading to the kinetic balance prescription.\cite{Stanton1984,dyall:jpb1984,Dyall_CPL1990} However, it is not possible to take this limit for the positive- and negative-energy solutions of the Dirac equation at the same time. Since the focus in chemistry is definitely on the positive-energy solutions, the relativistic energy scale is aligned with the non-relativistic one through the substitution $E\rightarrow E-m_e c^2$, whereupon the limit is taken as
\begin{equation}\label{eq:rkb}
\lim_{c\rightarrow\infty}c\psi^S = \lim_{c\rightarrow\infty}\frac{1}{2m_e}\left[1+\frac {E-V}{2m_e c^2}\right]\left(\boldsymbol{\sigma}\cdot\mathbf{p}\right)\psi^L=
\frac{1}{2m_e}\left(\boldsymbol{\sigma}\cdot\mathbf{p}\right)\psi^L.
\end{equation}

In practice one may choose between 1- and 2-component basis functions for 4-component relativistic molecular calculations. The latter choice allows the straightforward realization of \textit{restricted} kinetic balance,\cite{Dyall_CPL1990} hence a 1:1 ratio of large and small component basis functions, but requires on the other hand a dedicated integration module. In DIRAC we opted\cite{Saue:MP1997} for Cartesian GTOs
\begin{equation}\label{eq:cGTO}
G^{\alpha}_{ijk}(\mathbf{r})=Nx^i y^j z^k\exp[-\alpha r^2];\quad i+j+k=\ell
\end{equation}
since this gave immediate access to integrals of the HERMIT integral module,\cite{prog:hermit} where the extensive menu of one-electron integrals boosted functionality in terms of molecular properties.

DIRAC provides a library of Gaussian basis sets. The main basis sets available are those of Dyall and coworkers,\cite{dyall:tca1998,dyall:tca2002,dyall:tca2004,dyall:tca2006,dyall:tca2007a,dyall:tca2007b,dyall:jpca2009,dyall:tca2010a,dyall:tca2010b,dyall:tca2011,dyall:tca2012a,dyall:tca2012b,dyall:tca2016} which cover all elements from H to Og at the double-zeta, triple-zeta, and quadruple-zeta level of accuracy. They include functions for electron correlation for valence, outer core and inner core, as well as diffuse functions, in the style of the Dunning correlation-consistent basis sets.\cite{dunning:jcp1989}

\subsection{SCF module}
The SCF module has some unique features that will
be described in the following. In matrix form the HF/KS equations read
\begin{equation}
  F\mathbf{c}=S\mathbf{c}\varepsilon
\end{equation}
where $F$ and $S$ are the Fock/KS and overlap matrices, respectively, and $\mathbf{c}$ refers to expansion coefficients. Before diagonalization the equations are transformed to an orthonormal
basis
\begin{equation}\label{eq:ortho}
  \tilde{F}\tilde{\mathbf{c}}=\tilde{\mathbf{c}}\varepsilon;\quad \tilde{F}=V^{\dagger}FV;\quad \mathbf{c}=V\tilde{\mathbf{c}};\quad V^{\dagger}SV=I.
\end{equation}
As a simple example we may take the Dirac equation in a finite basis,
\begin{equation}
    \left[\begin{array}{cc}
V^{LL} & c\Pi^{LS}\\
c\Pi^{SL} & V^{SS}-2m_{e}c^{2}S^{SS}
\end{array}\right]\left[\begin{array}{c}
\mathbf{c}^{L}\\
\mathbf{c}^{S}
\end{array}\right]=\left[\begin{array}{cc}
S^{LL} & 0\\
0 & S^{SS}
\end{array}\right]\left[\begin{array}{c}
\mathbf{c}^{L}\\
\mathbf{c}^{S}
\end{array}\right]E.
\end{equation}
After orthonormalization it reads
\begin{equation}
\left[\begin{array}{cc}
\tilde{V}^{LL} & c\tilde{\Pi}^{LS}\\
c\tilde{\Pi}^{SL} & \tilde{V}^{SS}-2m_{e}c^{2}\tilde{I}^{SS}
\end{array}\right]\left[\begin{array}{c}
\tilde{\mathbf{c}}^{L}\\
\tilde{\mathbf{c}}^{S}
\end{array}\right]=\left[\begin{array}{cc}
\tilde{I}^{LL} & 0\\
0 & \tilde{I}^{SS}
\end{array}\right]\left[\begin{array}{c}
\tilde{\mathbf{c}}^{L}\\
\tilde{\mathbf{c}}^{S}
\end{array}\right]E    
\end{equation}
DIRAC employs canonical orthonormalization\cite{Lowdin:AQC1970} which allows the elimination of linear dependencies. However, the orthonormalization step is overloaded:
\begin{enumerate}

\item\textit{Elimination and freezing of orbitals:} 
DIRAC allows the elimination and freezing of orbitals. Such orbitals are provided by the user in the form of one or more coefficient files. This part of the code uses the machinery of the projection analysis discussed in Section \ref{visualization}. The selected orbitals can therefore be expressed either in the full molecular basis or in the basis set of some chosen (atomic) fragment. They are eliminated by transforming them to the orthonormal basis and projecting them out of the transformation matrix $V$.  They may instead be frozen by putting them back in the appropriate position when back-transforming coefficients to the starting AO basis.

An example of the use of elimination of orbitals is a study of the effect of the lanthanide contraction on the spectroscopic constants of the CsAu molecule.\cite{Fossgaard2003b} Inspired by an atomic study by Bagus and co-workers,\cite{bagus:pseudoatom} the precalculated \textit{4f}-orbitals of the gold atom were imported into a molecular calculation and eliminated. At the same time the gold nuclear charge was reduced by 14 units, thus generating a pseudo-gold atom unaffected by the lanthanide contraction. An example of the freezing of orbitals is the study of the effect of the freezing of oxygen \textit{2s}-orbitals on the electronic and molecular structure of the water molecule.\cite{Dubillard2006}

\item \textit{Cartesian-to-spherical transformation:} As already mentioned, at the integral level, DIRAC employs Cartesian Gaussian-type orbitals (GTOs), Eq.~\eqref{eq:cGTO}. One would perhaps rather have expected the use of the more economical spherical GTOs
\begin{equation}
G^{\alpha}_{\ell m}(\mathbf{r})=Nr^{\ell}\exp[-\alpha r^2]Y_{\ell m}(\theta,\phi).    
\end{equation}
However, this is precluded by the kinetic balance prescription. In the atomic case, the nonrelativistic limit of the coupling between the large and small radial functions reads
\begin{equation}
\lim_{c\rightarrow\infty}cQ_{\kappa}
=\frac{1}{2m_e}\left(\partial_{r}+\frac{\kappa}{r}\right)P_{\kappa},
\end{equation}
which in the present case implies
\begin{equation*}
P_{\kappa}(r)=Nr^{\ell-1}\exp[-\alpha r^2]\quad\Rightarrow\quad
Q_{\kappa}(r)=N\left\{(\kappa+\ell-1)r^{\ell-2}-2\alpha r^{\ell}\right\}\exp[-\alpha r^2].
\end{equation*}
Rather than implementing the transformation to the non-standard radial part of the small component spherical GTOs at the integral level, we have embedded it in the transformation to the orthonormal basis.

\item \textit{Restricted kinetic balance:} The use of scalar basis functions only allows \textit{unrestricted} kinetic balance, where the small component basis functions are generated as derivatives of the large component ones, but not in the fixed 2-component linear combination of Eq.~\eqref{eq:rkb}. This leads to the curious situation that the small component basis is represented by more functions than the large component one, e.g. a single large component \textit{s}-function generates three small component \textit{p}-functions. In DIRAC we do, however, recover RKB in the orthonormalization step. In the first version, RKB was obtained by noting that the extra small component basis functions mean that there will be solutions of the Dirac equation with zero large components. In the free-particle case these solutions will have energy $-2m_e c^2$. RKB was therefore realized by diagonalizing the free-particle Dirac equation in orthonormal basis, then identifying and eliminating (as described above) these unphysical solutions. 

It was later realized that RKB could be achieved in a simpler manner by embedding the transformation to the modified Dirac equation\cite{dyall:jcp1994,visscher:jcp2000}
\begin{equation}\label{eq:modDir}
\begin{array}{l}
Q=\left[\begin{array}{cc}
\tilde{I}^{LL} & 0\\
0 & \frac{1}{2m_{e}c}\tilde{\Pi}^{SL}
\end{array}  
\right]\quad\Rightarrow
\\ \\
\left[\begin{array}{cc}
\tilde{V}^{LL} & \frac{1}{2m_{e}}\tilde{T}^{LL}\\
\frac{1}{2m_{e}}\tilde{T}^{LL} & \frac{1}{4m^{2}c^{2}}\tilde{W}^{LL}-\frac{1}{2m_{e}}\tilde{T}^{LL}
\end{array}\right]\left[\begin{array}{c}
\tilde{\mathbf{c}}^{L\prime}\\
\tilde{\mathbf{c}}^{S\prime}
\end{array}\right]=\left[\begin{array}{cc}
\tilde{I}^{LL} & 0\\
0 & \frac{1}{4m^{2}c^{2}}\tilde{T}^{LL}
\end{array}\right]\left[\begin{array}{c}
\tilde{\mathbf{c}}^{L\prime}\\
\tilde{\mathbf{c}}^{S\prime}
\end{array}\right],
\end{array}
\end{equation}
where
\begin{eqnarray}
\tilde{T}^{LL}_{\mu\nu} & = &\displaystyle \frac{1}{2m_{e}}\sum_{\gamma}\tilde{\Pi}_{\mu\gamma}^{LS}\tilde{\Pi}_{\gamma\nu}^{SL}=\langle\tilde{\chi}_\mu|\frac{p^2}{2m_e}|\tilde{\chi}_{\nu}\rangle\label{eq:Tll}\\
\tilde{W}^{LL}_{\mu\nu} & = &\displaystyle \sum_{\gamma\delta}\tilde{\Pi}^{LS}_{\mu\gamma}\tilde{V}^{SS}_{\gamma\delta}\tilde{\Pi}^{SL}_{\delta\nu}=\langle\tilde{\chi}_{\mu}|(\boldsymbol{\sigma}\cdot\mathbf{p})V(\boldsymbol{\sigma}\cdot\mathbf{p})|\tilde{\chi}_\nu\rangle\label{eq:Wll},
\end{eqnarray}
where the latter equalities follow from kinetic balance.\cite{dyall:jpb1984} 
The metric on the right-hand side of Eq.~\eqref{eq:modDir} indicates a non-orthonormal basis. A second canonical orthonormalization transformation $\tilde{V}$ is therefore introduced, so that the total transformation, done in a single step, reads $VQ\tilde{V}$.

\item \textit{Elimination of spin-orbit interaction}: As shown by Dyall,\cite{dyall:jcp1994} transformation to the modified Dirac equation allows a separation of the spin-free and spin-dependent terms. In the quaternion symmetry scheme of DIRAC, we obtain such a separation by simply deleting the quaternion imaginary parts of Fock matrices in the orthonormal basis.\cite{visscher:jcp2000}
\item \textit{X2C transformation}: The transformation to the eXact 2-Component relativistic (X2C) Hamiltonian is carried out starting from the modified Dirac equation in the orthonormal basis. Working with a \textit{unit} metric greatly simplifies the transformation.\cite{liu:jcp2009}
\item \textit{Supersymmetry}: At the integral level, basis functions are adapted to symmetries of $D_{2h}$ and subgroups. However, for linear systems, we obtain a blocking of Fock matrices in the orthonormal basis on the $m_j$ quantum number\cite{Visscher:linsym} by diagonalizing the matrix of the $\hat{j}_z$ operator in the orthonormal basis and performing the substitution $V\rightarrow VU_m$, where $U_m$ are the eigenvectors ordered on $m_j$. This provides significant computational savings, in particular at the correlated level. Recently we have implemented atomic supersymmetry, such that the Fock matrix gets blocked on $(\kappa, m_j)$ quantum numbers (to appear in DIRAC20).\cite{Sunaga:atsym}
\end{enumerate}

\subsection{Symmetry considerations}\label{subsec:symmetry-considerations}
The DIRAC code can handle symmetries corresponding to $D_{2h}$ and subgroups (denoted binary groups) as well as linear (and atomic) supersymmetry.  At the SCF level DIRAC employs a unique quaternion symmetry scheme which combines time reversal and spatial symmetry.\cite{Saue:Jensen:JCP1999} A particularity of this scheme is that symmetry reductions due to spatial symmetry are translated into a reduction of algebra, from quaternion down to complex and possibly real algebra. This leads to a classification of the binary groups as:
\begin{itemize}
    \item Quaternion groups: $C_1$, $C_i$
    \item Complex groups: $C_2$, $C_s$, $C_{2h}$
    \item Real groups: $D_2$, $C_{2v}$, $D_{2h}$
\end{itemize}
At the SCF level DIRAC works with the irreducible \textit{co-representations} obtained by combining the above spatial symmetry groups with time reversal symmetry.\cite{Saue:Jensen:JCP1999} 
A source of confusion for DIRAC users is that occupations are given for each irreducible co-representation at the SCF level. However, one can show that starting from the binary groups, there are at most two irreducible co-representations, distinguished by parity. This means in practice that a single occupation number is expected for systems without inversion symmetry, whereas occupations for \textit{gerade} and \textit{ungerade} symmetries are given separately otherwise.

At the correlated level, the highest Abelian subgroup of the point group under consideration is used. For the point groups implemented this leads to the following group chains:
\begin{itemize}
    \item $D_2,C_{2v}\rightarrow C_2$
    \item $D_{2h}\rightarrow C_{2h}$
    \item $C_{\infty v}\rightarrow C_{64v}$
    \item $D_{\infty h}\rightarrow D_{32h}$.
\end{itemize}
 
The linear groups $D_{\infty h}$ and $C_{\infty v}$ are special as the number of finite Abelian subgroups that can be used to characterize orbitals is infinite. In practice we map these groups to a 64-dimensional subgroup, which is more than sufficient to benefit from symmetry blocking in the handling of matrices and integrals and to identify the symmetry character of orbitals and wave functions. The group chain approach,\cite{Nieuwpoort_1961} in which each orbital transforms according to the irreps of the Abelian subgroup as well as a higher, non-Abelian group has as advantage that the defining elements of the second quantized Hamiltonian of Eq.~\eqref{eq:secham} are real for the real groups, even though an Abelian complex group is used at the correlated level. The transition between the quaternion algebra used at the SCF level and the complex or real algebra used in the correlation modules is made in the AO-to-MO transformation which generates transformed integrals in quaternion format, after which they are expressed and stored in a complex (or real) form.\cite{Visscher:2002p592}

The use of real instead of complex algebra gives a fourfold speed-up for floating point multiplications. In RELCCSD one generic algorithm is used for all implemented point groups, with the toggling between complex or real multiplications hidden inside a wrapper for matrix multiplications. The LUCIAREL kernel has distinct implementations for real-valued and complex-valued Abelian double point groups.\cite{fleig_gasmcscf} For linear molecules \cite{Denis2015} and atoms \cite{FleigJung_JHEP2018} axial symmetry is useful and implemented. \cite{knecht_u2} For linear groups an isomorphic mapping between total angular momentum projection (along the distinguished axis) and group irreducible representation is possible for all practically occurring angular momenta using the 64-dimensional subgroups defined above.

\section{Conclusions}
DIRAC is one of the earliest codes for 4-component relativistic molecular calculations and the very first to feature exact 2-component (X2C) relativistic calculations.\cite{ilias:cpl2005} It is not the fastest such code around, but is presently hard to beat in terms of functionality. This stems in part from the fact that the code has been written with generality in mind. There is a wide range of Hamiltonians, and most program modules are available for all of them. The SCF module allows Kramers-restricted HF and KS calculations using an innovative symmetry scheme based on quaternion algebra. In some situations, though, for instance in SCF calculations of magnetic properties, unrestricted calculations are desirable in order to capture spin polarization. 

A number of \textit{molecular} properties, such as electric field gradients,\cite{Visscher_JCP1998} parity-violation in chiral molecules,\cite{Laerdahl:PRA1999} nuclear spin-rotation constants\cite{Aucar:REHE2012,Aucar:JCP2013} and rotational g-tensors\cite{Aucar:JCP2014} were first studied in a 4-component relativistic framework with DIRAC. The freedom of users to define their own properties combined with the availability of properties up to third order means that there are many new properties waiting to be explored. Such properties may be further analyzed through the powerful visualization module.

Another strength of DIRAC is the large selection of wave function-based correlation methods, including MRCI, CCSD(T), FSCCSD, EOM-CCSD, ADC and MCSCF. As already mentioned, the latter allowed a detailed study of the emblematic U$_2$ molecule, demonstrating that spin-orbit interaction reduces the bond order from five\cite{roos:U2} to four.\cite{knecht_u2} 
Methods implemented in DIRAC that account for more dynamic correlation have, combined with experiment, provided reference values for properties such as nuclear quadrupole moments,\cite{vanStralen:2002p111, VanStralen:2003p103} hyperfine structure  constants\cite{haase2020hyperfine} and M{\"o}ssbauer isomer shifts.\cite{Zelovich_MP2017} DIRAC also provides theoretical input for spectroscopic tests of fundamental physics, both within the Standard Model of elementary particles \cite{Hao_PhysRevA.98.032510} as well as tests of Beyond Standard Model (BSM) theories which give rise to electric dipole moments of fermions.
\cite{Denis2015,Denis:2019bx,Skripnikov_ThO_JCP2016}

In recent years, DIRAC has been extended to include several models for large environments: PCM, PE and FDE, which opens new perspectives. For instance, recently EOM-CC was combined with FDE to calculate ionization energies of halide ions in droplets modelled by 50 water molecules.\cite{halides-water-Bouchafra-PRL2019} 


We believe it is safe to say that DIRAC is a reference in the domain of 2- and 4-component relativistic molecular calculations and that it will remain so in the foreseeable future. In 2015 DIRAC was one of 13 scientific software suites chosen for adaption to the SUMMIT supercomputer \cite{Straatsma_CAAR2020} at the Oak Ridge Leadership Computing Facility (OLCF). As of November 2019, SUMMIT was the world's fastest supercomputer, and DIRAC production runs are currently being carried out on this machine.

\begin{acknowledgments}
TS would like to thanks his former advisors Knut F{\ae}gri jr. and the late Odd Gropen for putting him on an exciting track.

LV acknowledges support of the Dutch Research Council (NWO) for this research via various programs. He also likes to thank his former advisors Patrick Aerts and Wim Nieuwpoort for introducing him to the wonderful world of relativistic quantum chemistry.

ASPG acknowledges support from the CNRS Institute of Physics (INP), PIA ANR project CaPPA(ANR-11-LABX-0005-01), I-SITE ULNE project OVERSEE (ANR-16-IDEX-0004), the French Ministry of Higher Education and Research, region Hauts de France council and European Regional Development Fund (ERDF) project CPER CLIMIBIO.

MI acknowledges the support of the Slovak Research and Development Agency and the Scientific Grant Agency, APVV-15-0105 and VEGA  1/0562/20, respectively. This research used resources of a High Performance Computing Center of the Matej Bel University in Banska Bystrica using the HPC infrastructure acquired in projects ITMS 26230120002 and 26210120002 (Slovak infrastructure for high performance computing) supported by the Research and Development Operational Programme funded by the ERDF. 

RDR acknowledges partial support by the Research Council of Norway through its
Centres of Excellence scheme, project number 262695 and through its Mobility
Grant scheme, project number 261873.

MO acknowledges support of the Polish National Science Centre (2016/23/D/ST4/ 03217).

IAA acknowledges support from CONICET by grant PIP 112-20130100361 and FONCYT by grant PICT 2016-2936.

JMHO acknowledges financial support from the Research Council of Norway through its Centres of Excellence scheme (Project ID: 262695).

AS acknowledges financial support from Japan Society for the Promotion of Science (JSPS) KAKENHI Grants No. 17J02767, and JSPS Overseas Challenge Program for Young Researchers Grants No. 201880193.

SK would like to thank Markus Reiher (ETH Z\"urich) for his continuous support throughout his time at ETH Z\"urich. 

\end{acknowledgments}

\bibliography{article}

\end{document}